\documentclass[twoside,twocolumn,9pt]{article}
\usepackage{extsizes}
\usepackage[super,sort&compress,comma]{natbib} 
\usepackage[version=3]{mhchem}
\usepackage[left=1.5cm, right=1.5cm, top=1.785cm, bottom=2.0cm]{geometry}
\usepackage{balance}
\usepackage{mathptmx}
\usepackage{sectsty}
\usepackage{graphicx} 
\usepackage{lastpage}
\usepackage[format=plain,justification=justified,singlelinecheck=false,font={stretch=1.125,small,sf},labelfont=bf,labelsep=space]{caption}
\usepackage{float}
\usepackage{fancyhdr}
\usepackage{fnpos}
\usepackage[english]{babel}
\addto{\captionsenglish}{%
  
}
\usepackage{array}
\usepackage{droidsans}
\usepackage{charter}
\usepackage[T1]{fontenc}
\usepackage[usenames,dvipsnames]{xcolor}
\usepackage{setspace}
\usepackage[compact]{titlesec}
\usepackage{hyperref}

\usepackage{amsmath, amsthm, amssymb, amsfonts}
\usepackage{bm}
\usepackage[normalem]{ulem}
\usepackage{longtable}
\usepackage{multirow}
\usepackage{booktabs}   
\usepackage{listings}
\usepackage[usenames,dvipsnames]{xcolor}
\usepackage{subfiles}
\usepackage{pdflscape}
\usepackage{siunitx}

\usepackage{epstopdf}

\usepackage{float}

\definecolor{cream}{RGB}{222,217,201}

\DeclareRobustCommand{\OH}{\ensuremath{\mathrm{HO}^{\bullet}}}
\DeclareRobustCommand{\Hdot}{\ensuremath{\mathrm{H}^{\bullet}}}
\DeclareRobustCommand{\eaq}{\ensuremath{e^-_{\mathrm{aq}}}} 
\DeclareRobustCommand{\HtwoOtwo}{\ensuremath{\mathrm{H_2O_2}}}
\DeclareRobustCommand{\HthreeOplus}{\ensuremath{\mathrm{H_3O^+}}}
\DeclareRobustCommand{\OHminus}{\ensuremath{\mathrm{HO^-}}} 
\DeclareRobustCommand{\Htwo}{\ensuremath{\mathrm{H_2}}}
\DeclareRobustCommand{\HtwoOplus}{\ensuremath{\mathrm{H_2O^+}}}
\DeclareRobustCommand{\HtwoO}{\ensuremath{\mathrm{H_2O}}}
\DeclareRobustCommand{\HtwoOstar}{\ensuremath{\mathrm{H_2O^*}}}
\DeclareRobustCommand{\esol}{\ensuremath{e^-_{\mathrm{aq}}}}

\begin{document}

\pagestyle{fancy}
\thispagestyle{plain}
\fancypagestyle{plain}{
\renewcommand{\headrulewidth}{0pt}
}

\makeFNbottom
\makeatletter
\renewcommand\LARGE{\@setfontsize\LARGE{15pt}{17}}
\renewcommand\Large{\@setfontsize\Large{12pt}{14}}
\renewcommand\large{\@setfontsize\large{10pt}{12}}
\renewcommand\footnotesize{\@setfontsize\footnotesize{7pt}{10}}
\makeatother

\renewcommand{\thefootnote}{\fnsymbol{footnote}}
\renewcommand\footnoterule{\vspace*{1pt}%
\color{cream}\hrule width 3.5in height 0.4pt \color{black}\vspace*{5pt}} 
\setcounter{secnumdepth}{5}

\makeatletter 
\renewcommand\@biblabel[1]{#1}            
\renewcommand\@makefntext[1]%
{\noindent\makebox[0pt][r]{\@thefnmark\,}#1}
\makeatother 
\renewcommand{\figurename}{\small{Fig.}~}
\sectionfont{\sffamily\Large}
\subsectionfont{\normalsize}
\subsubsectionfont{\bf}
\setstretch{1.125} 
\setlength{\skip\footins}{0.8cm}
\setlength{\footnotesep}{0.25cm}
\setlength{\jot}{10pt}
\titlespacing*{\section}{0pt}{4pt}{4pt}
\titlespacing*{\subsection}{0pt}{15pt}{1pt}

\fancyfoot{}
\fancyfoot[LO,RE]{\vspace{-7.1pt}\includegraphics[height=9pt]{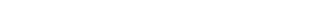}}
\fancyfoot[CO]{\vspace{-7.1pt}\hspace{13.2cm}\includegraphics{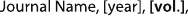}}
\fancyfoot[CE]{\vspace{-7.2pt}\hspace{-14.2cm}\includegraphics{head_foot/RF}}
\fancyfoot[RO]{\footnotesize{\sffamily{1--\pageref{LastPage} ~\textbar  \hspace{2pt}\thepage}}}
\fancyfoot[LE]{\footnotesize{\sffamily{\thepage~\textbar\hspace{3.45cm} 1--\pageref{LastPage}}}}
\fancyhead{}
\renewcommand{\headrulewidth}{0pt} 
\renewcommand{\footrulewidth}{0pt}
\setlength{\arrayrulewidth}{1pt}
\setlength{\columnsep}{6.5mm}
\setlength\bibsep{1pt}

\makeatletter 
\newlength{\figrulesep} 
\setlength{\figrulesep}{0.5\textfloatsep} 

\newcommand{\topfigrule}{\vspace*{-1pt}%
\noindent{\color{cream}\rule[-\figrulesep]{\columnwidth}{1.5pt}} }

\newcommand{\botfigrule}{\vspace*{-2pt}%
\noindent{\color{cream}\rule[\figrulesep]{\columnwidth}{1.5pt}} }

\newcommand{\dblfigrule}{\vspace*{-1pt}%
\noindent{\color{cream}\rule[-\figrulesep]{\textwidth}{1.5pt}} }

\makeatother

\twocolumn[
  \begin{@twocolumnfalse}
{\includegraphics[height=30pt]{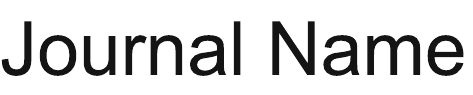}\hfill\raisebox{0pt}[0pt][0pt]{\includegraphics[height=55pt]{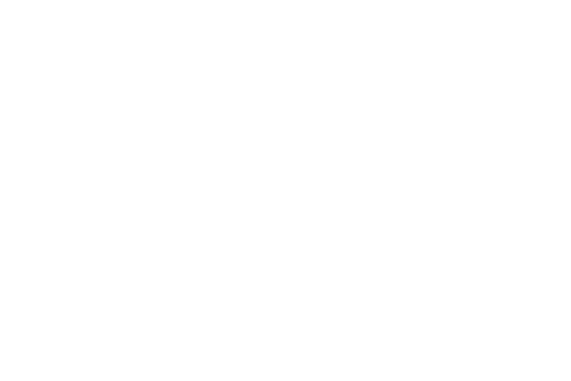}}\\[1ex]
\includegraphics[width=18.5cm]{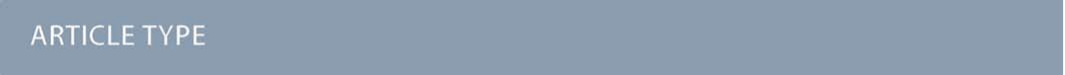}}\par
\vspace{1em}
\sffamily
\begin{tabular}{m{4.5cm} p{13.5cm} }

\includegraphics{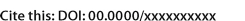} & \noindent\LARGE{\textbf{Simulation of Radiation Chemistry by a One-Shot Hybrid Continuum / Monte Carlo Method}} \\
\vspace{0.3cm} & \vspace{0.3cm} \\

 & \noindent\large{Charlie Fynn Perkins,\textit{$^{a}$} Marcus Webb,\textit{$^{b}$} and Fred J. Currell\textit{$^{c,d}$}} \\

\includegraphics{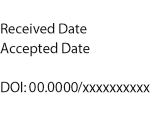} & \noindent\normalsize{Understanding the spatio-temporal evolution of radiolytic species created by high-energy electrons in water underpins key applications from radiotherapy and nuclear safety to environmental processing and electron microscopy. Here, using the Manchester Inhomogeneous Radiation Chemistry by Linear Expansions (MIRaCLE) toolkit, we introduce and benchmark a novel approach to simulating these processes.  Although the initial conditions are determined stochastically, the subsequent time evolution is calculated deterministically using a continuum representation, derived from those initial conditions. This hybrid approach essentially averages over many chemistry `trajectories' simultaneously, often converging to the $1$\% level in one shot, not requiring multiple runs. We demonstrate this new approach through the calculation of time-dependent $G$-values for \eaq, \OH \; and other radiolytic products, including at unprecedented dose rates where calculations which would take years with a conventional Monte Carlo approach can be performed in mere hours on a commercial laptop. We demonstrate that the main artifact of continuum modelling can be mitigated by a correction term. These results establish MIRaCLE as a flexible and efficient platform for modelling long-timescale radiolysis, providing a bridge between Monte Carlo approaches and macroscopic reaction–diffusion schemes, with broad implications for radiation chemistry in medicine, energy, and materials science.} \\

\end{tabular}

 \end{@twocolumnfalse} \vspace{0.6cm}

  ]

\renewcommand*\rmdefault{bch}\normalfont\upshape
\rmfamily
\section*{}
\vspace{-1cm}


\footnotetext{\textit{$^{a}$~School of Mathematics, Statistics and Physics, University of Newcastle, Newcastle upon Tyne NE1 7RU, UK. E-mail: c.perkins2@newcastle.ac.uk}}
\footnotetext{\textit{$^{b}$~Department of Mathematics, The University of Manchester, Manchester M13 9PL, UK. E-mail: marcus.webb@manchester.ac.uk}}
\footnotetext{\textit{$^{c}$~Department of Chemistry, The University of Manchester, Manchester M13 9PL, UK}. E-mail: frederick.currell@manchester.ac.uk}
\footnotetext{\textit{$^{d}$~Dalton Cumbrian Facility, The University of Manchester, Cumbria CA24 3HA, UK}}

\footnotetext{\dag~Supplementary Information available. See DOI: 00.0000/00000000.}




\section{Introduction} \label{sec:intro}
Interactions of high-energy electrons with water is the key driver in radiolytic systems. Even if they are not the primary form of radiation under consideration, electrons are produced through ionising interactions. Hence, they are present by definition whenever ionising radiation is present in a system. Accordingly, understanding their properties is key to understanding radiation chemistry in any fluid medium subjected to ionising radiation. There are rich and complex dynamics involving many downstream processes, as described by \cite{jay-gerin_fundamentals_2025} and references therein. Understanding these dynamics underpins diverse applications: precision radiotherapy (including FLASH, where pulse structure perturbs spur chemistry)~\cite{wardman_flash_2020}; water-chemistry control and nuclear safety in fission and fusion systems~\cite{macdonald_radiolysis_2022,elliot_reaction_2009}; radioactive-waste management, hydrogen generation, and corrosion phenomena at interfaces~\cite{das_critical_2013}; environmental remediation and wastewater treatment by radiation processing~\cite{getoff_waterpollutants_1996,wojnarovits_radiation_2016}; food irradiation/sterilisation~\cite{woods_pikaev_1994}; medical dosimetry and standards (water as the reference phantom; beam-quality correction)~\cite{medin_beamquality_2006}; space and planetary aqueous radiochemistry~\cite{draganic_earthbeyond_1993}; and mitigating artefacts in \emph{in situ} liquid-phase electron microscopy~\cite{fritsch_liquidem_2025}. More generally, quantitative links from track-structure chemistry to biomolecular damage inform radiobiology and DNA-lesion chemistry~\cite{vonsonntag_dna_2006}.

Here, we demonstrate a novel hybrid stochastic continuum approach. A family of models is presented, which go from considering radiation fields as purely homogenous, to ones which fully differentiate between the various physico and physicochemical processes. Except for a purely homogenous model, these models use a set of initial conditions created stochastically through consideration of the physical processes taking place. From these initial conditions, a continuum representation of the reactive species present is derived and evolved deterministically through time, following a Partial Differential Equation (PDE). The family of models offers a trade-off between performance and accuracy with the most physically accurate descriptions delivering excellent agreement with pulsed radiolysis experiments and Monte Carlo simulations. We show that the approach described is scalable to very large numbers of electrons per simulation. 

Pleasingly, although the initial conditions are derived stochastically, for the simulations presented, we found that the results were very similar for different sets of initial conditions, i.e.~the simulation can often be done in `one shot'. We therefore refer to those that exhibit this behaviour and successfully predict the G-values as \textsc{COSMIC} (Continuum/One-Shot Monte Carlo Ionisation Chemistry) models.  Here we describe this novel approach and specifically its implementation within the MIRaCLE toolkit.

We use the term `vertex' throughout to denote a single site where ionisation or excitation has taken place and the term `reaction packet' to denote the continuum representation of the species created by that single event. These terms are distinct from the commonly used concept of a `spur' in radiation chemistry, since several vertices and the products emitted from them can constitute a spur.
\section{Method} \label{sec:method}

\subsection{MIRaCLE} \label{sec:method-MIRaCLE}

The Manchester Inhomogeneous Radiation Chemistry by Linear Expansion (MIRaCLE) package is a software toolkit for the simulation of chemical kinetics that is inhomogeneous in both space and time, currently in development at The University of Manchester \cite{bradshaw_new_2023,bradshaw2025spectral}. It is not yet available to the general public. While motivated by problems in radiation chemistry, the framework is general and can be applied to general chemical kinetics.

MIRaCLE implements a continuum modelling approach. The user specifies the geometry and the medium, the initial spatial distribution of species, their diffusion coefficients, the reaction network with associated rate constants, and any interactions at the interfaces. These are all specified in chemically transparent units; for instance, rate of adsorption onto a two-dimensional interface is entered in units of $[M/nm^2]/[M/nm^3]/ns = m/s$. From these inputs, MIRaCLE automatically generates the governing partial differential equations and boundary conditions.

The equations are solved numerically using adaptive spectral methods \cite{shen2011spectral}. The implementation allows automatic handling of spatial and temporal resolution, reducing the need for user specification of the parameters related to the numerical solution. In most cases, the user only specifies the chemistry, with no need to specify or understand these technical parameters at all. Simulation results can be queried at arbitrary positions and times using efficient high-order interpolation. MIRaCLE also provides access to observables and derived quantities, including spatially resolved concentrations, integrated species populations, and reaction yields. 

\subsection{Electron Models} \label{sec:method-full-radiolysis}
\begin{figure*}[h]
    \centering
    \includegraphics[
        width=0.9\linewidth,
        clip
    ]{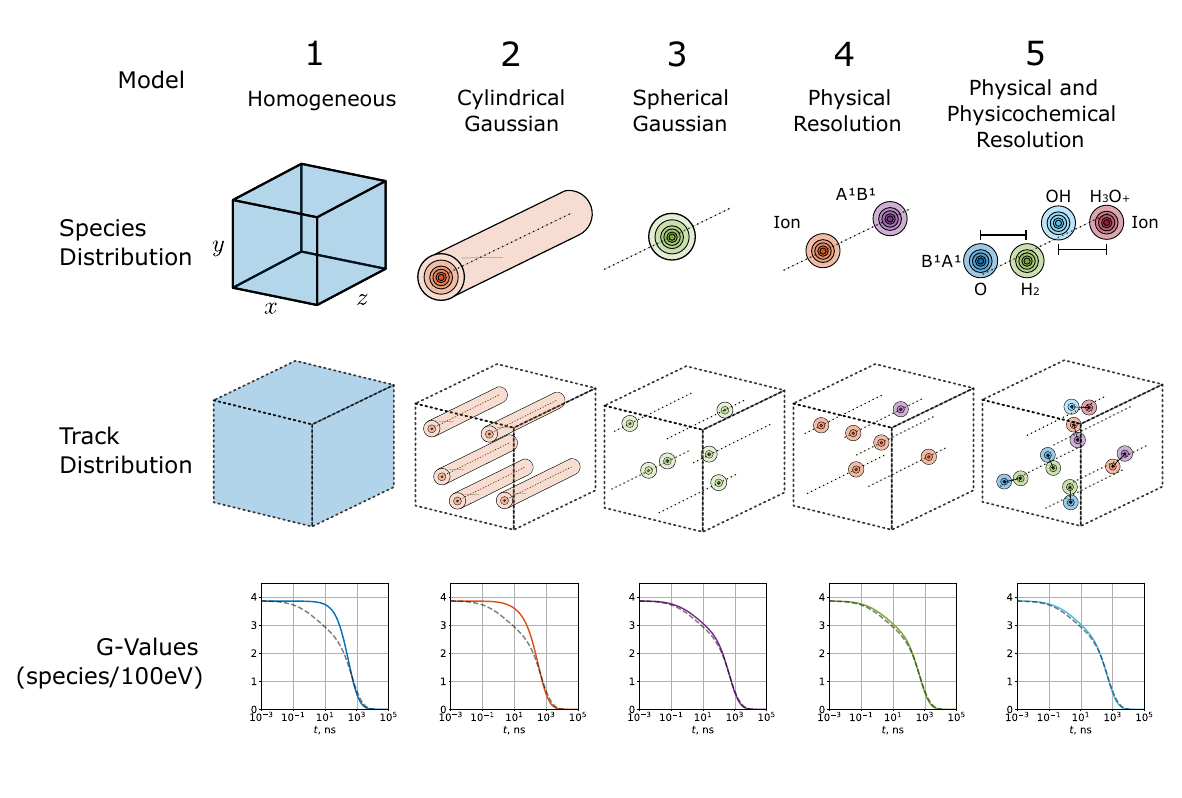}
    \caption{Illustrative overview of the models presented in this work, with the column number corresponding to the model numbers used in the main text. Rows contain, from top to bottom: the model number and name; an illustrative schematic of the reaction packets used to build the initial conditions, illustrative placement of several reaction packets used to build the full set of initial conditions and the computed G-value dynamics for solvated electrons.}
    \label{fig:inhom-illustrative}
\end{figure*}

 Five models of electron-induced water radiolysis were implemented, with varying levels of physical reality in terms of both spatial inhomogeneity and an increased resolution of the underlying mechanics. Fig.~\ref{fig:inhom-illustrative} illustrates each of these models. The models were benchmarked by simulating radiolysis in a $200\times200\times200\mathrm{nm^3}$ cube of water irradiated by $1 \; \mathrm{MeV}$ electrons, incident normal to the $x$-$y$ plane and distributed randomly in $x$-$y$. The conditions were chosen to match those of \cite{tran_geant4-dna_2021}; the absorbed dose for each simulation was $100 \; \mathrm{Gy}$ in the volume. The G-values for  $1 \; \mathrm{ps}$ yields were also taken from \cite{tran_geant4-dna_2021}, again to facilitate comparison. The maximum number of spectral modes per dimension used in MIRaCLE \cite{bradshaw_new_2023} was set 100 for each simulation and periodicity was enforced at domain boundaries. Simulations were performed from $1\;\mathrm{ps}$ to $100\;\mathrm{\mu s}$.

Model 1 treated the electron `beam' as homogenously entering the volume from one face. The concentration for each species was calculated using literature G-values, see Table \ref{tab:combined-physchem-data} in the Electronic Supplementary Information (ESI). Model 2 included the structure associated with the incidence of electrons on the $x$-$y$ face, so that radiolytic species were deposited as a series of stochastically placed cylindrical Gaussians, each with an axis parallel to the z-axis and running right across the simulation volume in the z-direction. Justification of these choices is given in Section \ref{sec:SI_Model2} of the ESI. The widths of the cylindrical Gaussians are given in Table \ref{tab:SI_widths} in the ESI.

Model 2 does not account for the stochastic nature of ionisation/excitation as the electrons pass through the water. In model 3, the cylindrical Gaussian distributions are replaced by a series of spherical Gaussian distributions centred on discrete collision vertices, set stochastically along the electron path, with interspacing being drawn from an exponential distribution. This more accurately models the distinct nature of point-like ionisation/excitation within the medium. Additionally, in this model, the solvated electron distribution used is a gamma distribution, where particles are distributed radially from the production site according to:
\begin{equation} \label{eqn:kreipl-gamma}
    f_{\mathrm{rad}}(r) = \frac{4}{r_0^2}r\exp\bigg(\frac{-2r}{r_0}\bigg),
\end{equation}
where $r_0$ is the mean thermalisation distance of the secondary electron at its energy, and $r$ is the radial distance from the vertex. Assuming that species are distributed isotropically, the corresponding volumetric probability density function (PDF) about a single vertex is
\begin{equation} \label{eqn:gamma-singularity}
    f_{\mathrm{vol}}(r) = \frac{1}{\pi r_0^2 r}\exp\bigg(\frac{-2r}{r_0}\bigg)
\end{equation}
which is singular at the origin. This singularity is undesirable as it suggests point-like behaviour of a diffuse particle, which is not only physically unrealistic but also computationally inconvenient. Therefore, eqn (\ref{eqn:gamma-singularity})  was regularised by hollowing out the centre of the distribution and enforcing a hard cut-off, such that the volumetric probability density $f_{\mathrm{vol}}(r)=0$ for some $r<r_{\mathrm{cut}}$. The normalised volumetric probability density is then
\begin{equation} \label{eqn:gamma-regularised}
    f_{\mathrm{vol}}(r) = \begin{cases}
        0 & r<r_{\mathrm{cut}}, \\
        \frac{r-r_{\mathrm{cut}}}{\pi r^2r_0^2}\exp\Big(\frac{-2(r-r_{\mathrm{cut}})}{r_0}\Big) & r\ge r_{\mathrm{cut}.}
    \end{cases}
\end{equation}
Provided it is sufficiently small, the choice of the cut-off radius is unimportant. Here it has been taken to be the  Van der Waal's radius \cite{halle_biomolecular_2003} for water, $r_\mathrm{cut}=0.14\;\mathrm{nm}$. The widths of the distributions for the chemical species are given in Table \ref{tab:SI_widths} in the ESI.

In model 4, collision vertices were placed in the same way as for model 3 but each was assigned to a different type of interaction (ionisation, A$^1$B$^1$ excitation, B$^1$A$^1$ excitation, Ryd+Dif); details are given in Section \ref{sec:SI_method-model4} of the ESI. We refer to this separation as \emph{physical differentiation} since the model differentiates between two physical processes - ionisation and excitation. Accordingly, there are different types of reaction packet.

Model 5 also includes the effect of dissociation processes taking place during the physicochemical stage. Due to proton-transfer followed by dissociation, pairs of species move apart, initially ballistically, before thermalising. This effect is most important in the placement of \OH \;and \Hdot \; distributions. Their centres were offset from each other in opposite directions such that they moved away from each other, with the direction for each dissociation event chosen to uniformly sample all possible directions. Justifications for this model and details of spherical Gaussian placement are given in Section \ref{sec:SI_method-model-5} of the ESI. Since this model additionally incorporates effects due to proton-transfer assisted dissociation, we refer to this as \emph{physicochemical differentiation}.

\begin{figure*}
    \centering
    \includegraphics[width=0.8\linewidth,
     trim=0 25 0 10, 
    clip
    ]{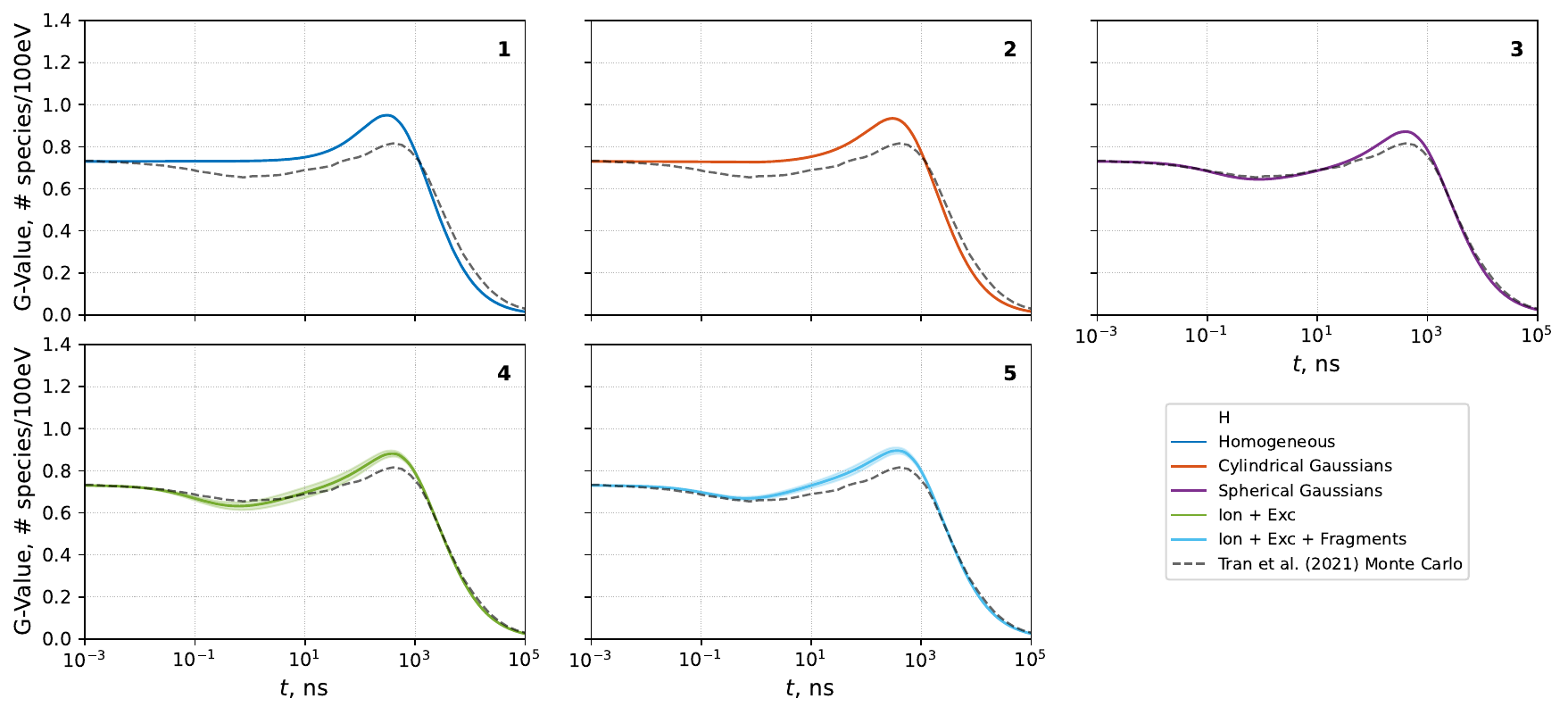}
    \caption{Comparison of hydrogen radical time dependent G-values across all models (coloured lines, panels 1 to 5) with the pure Monte Carlo study (greyed dashed line) by \cite{tran_geant4-dna_2021}, from $1$ps to $100\mu$s. Panel numbers correspond to model numbers.}
    \label{fig:models-compared-H-radical}
\end{figure*}

\subsection{Very High Dose Deposition} \label{sec:method-high-dose-rates}
To illustrate the ability of our new method to handle very high (quasi-instantaneous) dose depositions and to understand the scaling of computation time with the number of primary particles, we reran model 3 using deposited doses from $100 \;\mathrm{Gy}$ to $10,000\;\mathrm{Gy}$, with conditions otherwise being the same. Although this is a far higher 1 ps dose than can be achieved currently with electron beams, it has been used illustratively to demonstrate the potential of this method. It is noted that instantaneous electron dose rates of this order can be achieved in nanoparticle dose enhancement \cite{McMahon2011Biological} so it is valuable to demonstrate the method's capability to work in this regime.

\subsection{Pulse Radiolysis} \label{sec:method-pulse-radiolysis}
Model 3 was also used with conditions chosen to mimic picosecond pulsed radiolysis by $6-8\mathrm{MeV}$ electrons into \HtwoO \; \cite{wang_time-dependent_2018}. Once again, reduction in electron energy was neglected as is justified by stopping power data \cite{nist_star_2017}. In this experiment, the beam profile is inhomogeneous perpendicular to the beam direction over $\mathrm{mm}$ length scales, leading to variable current density. As such, one simulation was performed at a location within the beam which was chosen as a representative of the whole sample. Further details of the placement of this simulation is given in Supplementary Information \ref{sec:SI_method-pulse-radiolysis}.

\section{Results} \label{sec:results}

\subsection{Electron Models} \label{sec:results-full-radiolysis}

Time-dependent G-values for \Hdot \;across all models are shown in Fig.~ \ref{fig:models-compared-H-radical}. Each model was run 5 times with the average being displayed as the solid line. In a few cases, the standard deviation was greater than the line width used (e.g.~\Hdot \;for Models 4 and 5). Where  this occurs a line `envelope' has been used to illustrate the standard deviation. Fig.~\ref{fig:inhom-illustrative} shows a step-change in agreement going from Model 2 to Model 3. Interestingly, there is no further improvement as the extra physico and physicochemical differentiation is included. This highlights the remarkable effectiveness of Model 3 and is the reason it was used in subsequent studies.

Computational cost using an M2 Macbook Pro and $\chi$ goodness of fit values are presented for all models in Table \ref{tab:model-cost-chi}, as compared to Monte Carlo reference data \cite{tran_geant4-dna_2021}. Since there is a stochastic component to the simulation, which then can have a knock-on effect on the timings, 5 simulations were run in each case with the average and standard deviation being presented in the figures. The calculation of the $\chi$ values used the time-dependent mean of the five simulation runs for each model, compared against the corresponding reference data interpolated onto the same time grid. The expression for $\chi$ used is
\begin{equation} \label{eq:chi}
    \chi =\sqrt{\frac{1}{M N}\sum_{i=1}^{M}\sum_{j=1}^{N}\left(\frac{f_{ij} - g_i(t_j)}{\sigma_i}\right)^{2}},
\end{equation} 
where $f_{ij}$ is the mean simulated G-value for species $i$ at time $t_j$, $g_i(t_j)$ is the linearly interpolated reference value, $\sigma_i$ is the maximum reference G-value for species $i$, and $M$ and $N$ are the total number of species and time points respectively. Lower $\chi$ values represent closer agreement between our data and the reference data. As illustrated in Fig.~\ref{fig:inhom-illustrative} and  Fig.~\ref{fig:models-compared-H-radical}, Models 3 - 5 successfully reproduce the expected radiation chemistry, in one-shot, i.e.~they are the COSMIC models. Obtaining this agreement required no use of adjustable parameters or fine tuning on our part. All of the parameters used to specify the distributions of species have been taken from literature values.

\begin{table*}[tbp]
\centering
\setlength{\tabcolsep}{5pt} 
\resizebox{\textwidth}{!}{%
\begin{tabular}{lcccccccccc}
\toprule
\multirow{2}{*}{Model} & \multicolumn{2}{c}{Computational Cost (CPU-s)} & \multicolumn{8}{c}{$\chi$-values} \\
\cmidrule(lr){2-3}\cmidrule(lr){4-11}
 & Initial Conditions & Evolving the System & Total & \OH & \eaq & \HthreeOplus & \Hdot & \HtwoOtwo & \Htwo & \OHminus \\
\midrule
Model 1 & $4.11 \pm 0.06$ & $6.6 \pm 0.1$   & 0.156 & 0.188 & 0.101 & 0.075 & 0.083 & 0.289 & 0.045 & 0.164 \\
Model 2 & $10.5 \pm 0.6$  & $1470 \pm 55$   & 0.130 & 0.158 & 0.083 & 0.065 & 0.078 & 0.242 & 0.039 & 0.124 \\
Model 3 & $480 \pm 40$    & $3480 \pm 20$   & 0.064 & 0.045 & 0.017 & 0.012 & 0.026 & 0.144 & 0.050 & 0.051 \\
Model 4 & $270 \pm 20$    & $3510 \pm 60$   & 0.069 & 0.046 & 0.015 & 0.009 & 0.034 & 0.140 & 0.075 & 0.069 \\
Model 5 & $380 \pm 20$    & $3360 \pm 50$   & 0.075 & 0.046 & 0.014 & 0.008 & 0.046 & 0.164 & 0.062 & 0.066 \\
\bottomrule
\end{tabular}
}
\caption{Computational cost and $\chi$ goodness-of-fit values when compared to Monte Carlo  \cite{tran_geant4-dna_2021} calculations. Computational cost was measured with Julia's timing tools on a single CPU (M2 MacBook Pro). The total $\chi$ uses the scheme of Eq.~(\ref{eq:chi}); species-wise $\chi$ values use the same scheme without the sum over or normalisation by $M$.}
\label{tab:model-cost-chi}
\end{table*}

\begin{figure*}
    \centering
    \includegraphics[width=0.8\linewidth,
        trim=0 20 0 20, 
        ]{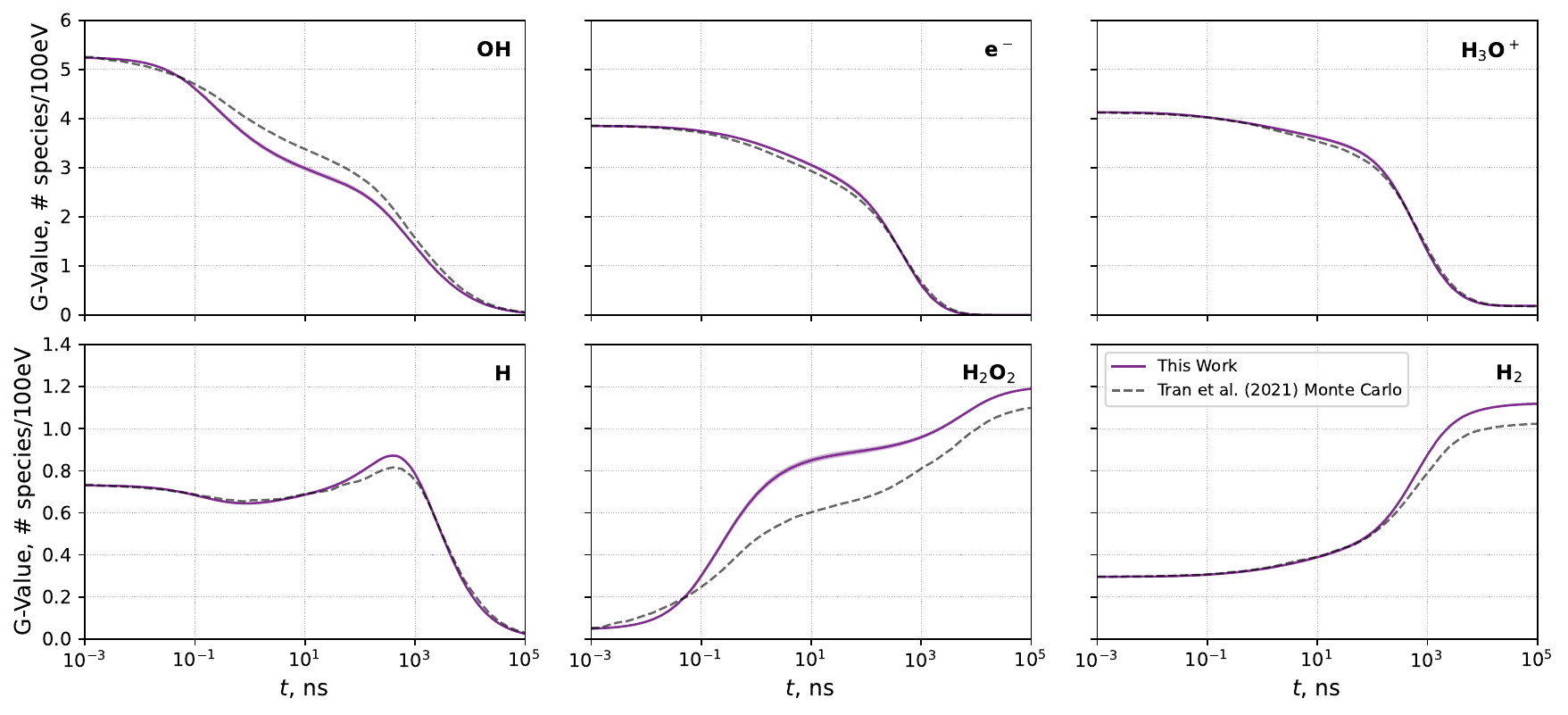}
    \caption{Time dependent G-values computed by Model 3 (simple spherical Gaussians) for 6 species (solid red line) and the Monte Carlo study (dashed grey line) of \cite{tran_geant4-dna_2021} from $1$ps to $100\mu$s.}
    \label{fig:model-3-g-vals}
\end{figure*}

\subsection{Higher Total Doses}

\begin{figure*}
    \centering
    \includegraphics[width=0.8\linewidth,
        trim=0 20 0 20, 
        ]{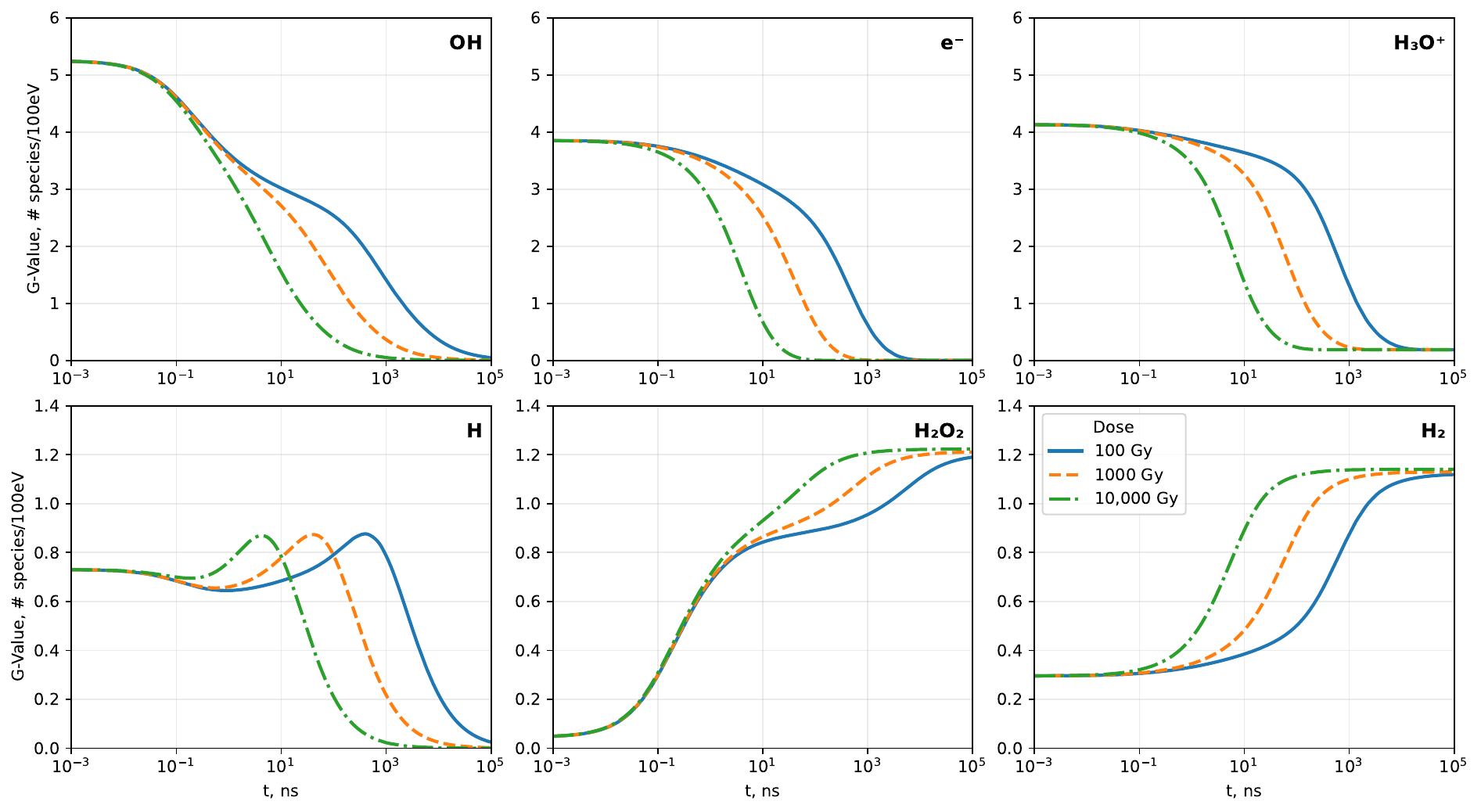}
    \caption{Time dependent G-values computed by Model 3 for dose depositions of 100 Gy, 1000 Gy and 10,000 Gy.}
    \label{fig:model-3-dose-rates}
\end{figure*}

Fig.~\ref{fig:model-3-dose-rates} shows the change in the predicted G-values as the dose deposited increases. As expected, higher dose leads to more reaction vertices which makes the initial conditions closer to homogeneous. As the reaction vertices from separate electron tracks increasingly overlap the rates of all of the reactions increase in a similar fashion since they are all second-order or pseudo second-order (see Table \ref{tab:combined-physchem-data} in the ESI). 

\subsection{Pulse Radiolysis} \label{sec:results-pulse-radiolysis}
\begin{figure}
    \centering
    \includegraphics[width=0.9\linewidth,
        trim=0 10 0 10, 
        ]{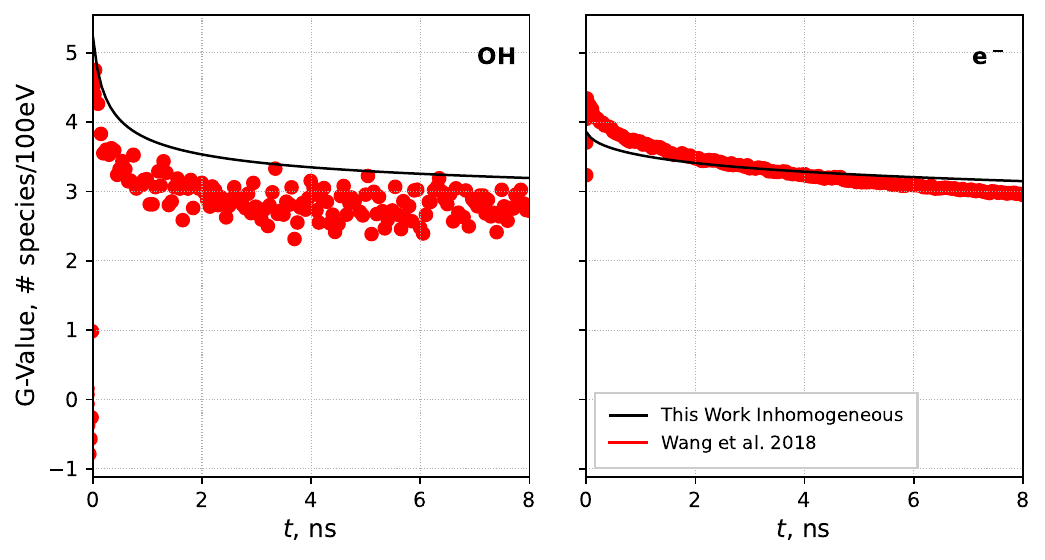}
    \caption{Time dependent G-values for \OH and \eaq for a study of pulse radiolysis. Simulated G-values using Model 3 (simple spherical Gaussians) are shown as a thick black line, and data points from the experiment performed by \cite{wang_time-dependent_2018} are shown as red dots, from $1$ ps to $8$ ns.}
    \label{fig:pulse-radiolysis}
\end{figure}

Unfortunately there are not many fast pulsed radiolysis studies against which these models can be compared. Time-dependent G-values for \OH \;and \eaq \;corresponding to the pulse radiolysis simulation described in Section \ref{sec:method-pulse-radiolysis} are presented in Fig.~\ref{fig:pulse-radiolysis}. The agreement is very good, especially when one recalls that there are no new adjustable parameters in the model. 

\section{Discussion} \label{sec:discussion}

\subsection{Electron Models and Pulsed Radiolysis} \label{sec:discussion-full-radiolysis}

Whilst Models 3--5 show good similarity with the Monte Carlo reference calculations \cite{tran_geant4-dna_2021}, it is clear that even these models show systematic variations for the G-values for \HtwoOtwo \;and \Htwo. These G-values are systematically too high, particularly after $10\, \mathrm{ns}$ (see Fig.~\ref{fig:model-3-g-vals}). Our continuum reaction-packet formulation systematically overestimates the G-values of both \(\mathrm{H_2}\) and \(\mathrm{H_2O_2}\) due to residual self-interactions within the reaction field. Reactive species originating from the same ionisation or excitation event are represented by overlapping probability densities rather than discrete particles. 
As a result,  radicals such as \Hdot \;and \OH \;have an enhanced likelihood of recombining with themselves, artificially increasing the rates of the homodimerization reactions 
$\Hdot + \Hdot \rightarrow \Htwo$ and $\OH + \OH \rightarrow \HtwoOtwo$. For example, an isolated reaction packet containing a single \Hdot\;should not be able to react with itself to produce \Htwo, however, this is possible within our approach.
This systematic bias, inherent to continuum representations, explains the consistent overproduction of \Htwo \;and \HtwoOtwo \;relative to particle-based simulations. 

\begin{figure}
    \centering
    \includegraphics[width=0.9\linewidth]{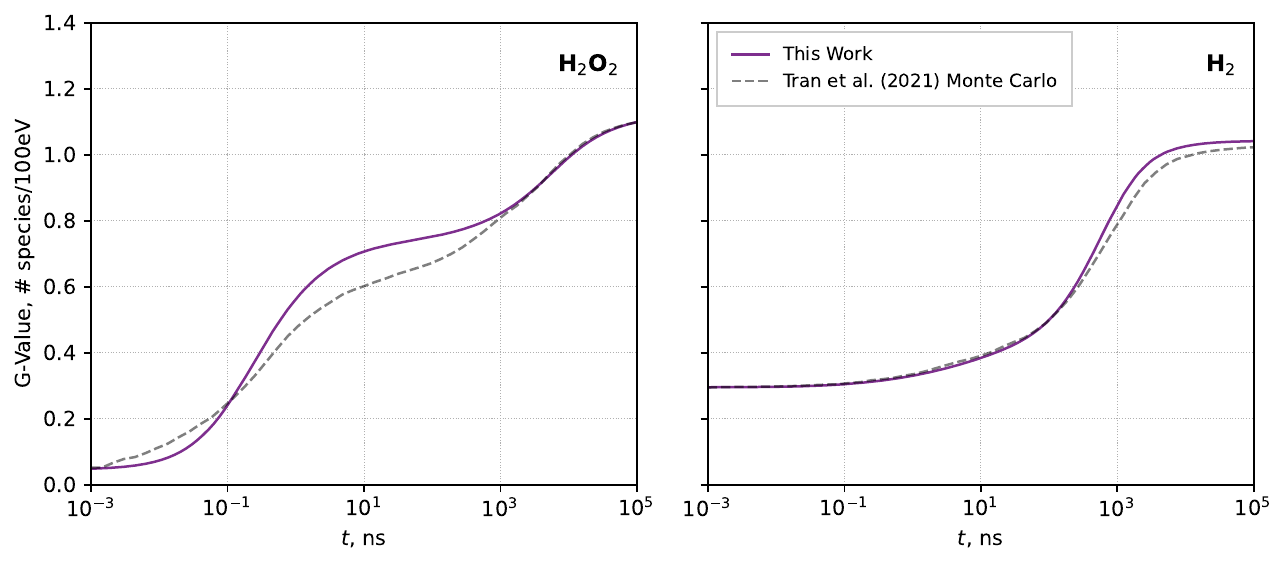}
    \caption{Time dependent G-values computed by Model 3 with algorithmically corrected rates (solid line) and comparison to the Monte Carlo reference (dashed line).}
    \label{fig:model-3-corrected-rates}
\end{figure}

For a given irradiation geometry and dose, it is possible to determine an analytical correction to the rate constants for these reactions, of the form:
\begin{equation}
\;k^{\mathrm{eff}}_{\mathrm{H\cdot+H\cdot\to H_2}}
= k_{\mathrm{HH}}\,
\frac{n_{\mathrm{H}}}{\,n_{\mathrm{H}}+J_{0,\mathrm{H}}(t_{\mathrm{eff}})\,},
\end{equation}
where $J_{0,\mathrm{H}}(t)$ is an analytical overlap integral describing the self-interaction, $n_{\mathrm{H}}$ is the vertex density and $k_{\mathrm{HH}}$  is the uncorrected rate constant (see Section \ref{sec:SI_modified-const} of the ESI). Entirely analogous expressions correct for the self-interaction for $\OH + \OH \rightarrow \HtwoOtwo$.
This single $k^{\mathrm{eff}}$ per channel corrects the asymptotic G-values for the specified dose and geometry, see Fig.~\ref{fig:model-3-corrected-rates}. 

We are currently developing a more generalised approach, applicable to any reaction scheme, which calculates the effect of these self interactions dynamically through a method that partitions each chemical species into artificial “colours” to prevent self-interactions, ensuring correct pairwise combinatorics while preserving the aggregate kinetics.

It is interesting to note that Models 4 and 5 exhibit significantly greater shot-to-shot variation, as indicated in Fig.~\ref{fig:models-compared-H-radical}. The origin of this variation is the comparatively low probability for excitation, particularly into the A$^1$B$^1$ state (see Table \ref{tab:SI_ballarini-branching-ratios} in the ESI). Since about 10\% of the vertices contain \Hdot, its population is susceptible to statistical variation in the initial-conditions' stochastic phase. This effect is even greater for the initial placement of \Htwo\;which occurs only 2.6\% of the time (see Table \ref{tab:SI_vertex-probabilities} in the ESI). In contrast, in Model 3, every reaction packet contains the correct proportion of \Hdot \;and \Htwo \;meaning it is far less subject to statistical variation. This is similar to the rare event problem encountered in Monte Carlo simulation and indicates that a large number of primary histories would be needed to achieve statistical convergence using that approach. From this viewpoint, whilst not physically so realistic, the sharing of \Hdot \;across many reaction packets in Model 3 can be viewed as somewhat analogous to splitting, in that rare but important events are represented many times with the effects then being weighted accordingly.

The pulse radiolysis simulation demonstrates the ability of our new approach in directly comparing to experimental results. No fine tuning of simulation parameters was undertaken by us to achieve these results --- all of the parameters being taken directly from the literature \cite{tran_geant4-dna_2021, kreipl_time_2009,wang_time-dependent_2018}, demonstrating MIRaCLE's capability in making predictions \emph{a priori}.

\subsection{High Dose Rate and Performance Scaling}

\begin{figure}
    \centering
    \includegraphics[width=0.8\linewidth]{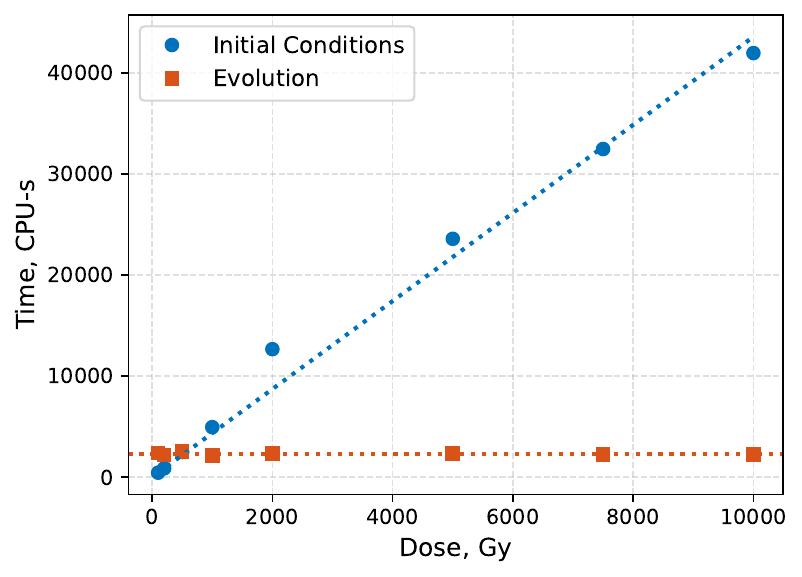}
    \caption{Computational time for a single core of an M2 Macbook Pro for setting up the initial conditions and for subsequently evolving the system as a function of dose deposited.}
    \label{fig:model-3-comp-times}
\end{figure}

The relative computational costs of two separate stages of the calculation are shown in Fig.~\ref{fig:model-3-comp-times}. The first cost concerns stochastically determining where the vertices should be placed and then using these vertices to create the initial continuum representations. This cost would be expected to scale linearly with dose, as is seen in Fig.~\ref{fig:model-3-comp-times}. The second computational cost shows the relative time for evolving the system from 1 ps to $100 \; \mu$s. Naively, this might be expected to be independent of time since MIRaCLE is handling the same amount of data in the continuum representation throughout.

It is difficult to make direct computational speed calculations, given the range of processing architectures, memory configurations, computer languages and compilers available. Never-the-less that data shown in Fig.~\ref{fig:model-3-dose-rates} and Fig.~\ref{fig:model-3-comp-times} illustrates a significant advantage of this new method. Any comparison to Monte Carlo step-by-step (SBS) of computational cost must necessarily be an approximate one and Tran et al.~\cite{tran_geant4-dna_2021} do not give the exact number of histories used. However, given that about 10\% of the histories will result in formation of a \Hdot \;radical, and this impacts the statistical convergence, to achieve statistical convergence to the 1\% level, at least $10^5$ primary electron histories would need to be considered. Since only 2.6\% of the vertices produce \Htwo, an order of magnitude more histories would be required to deliver statistical convergence for this species.

Camazzola et al.~\cite{Camazzola2023} quote a run time of 75 CPU-s per 500 electron tracks each with an energy of 500 keV, computing the chemical evolution for 10\,$\mu$s. Hence, to only compute out to 10\,$\mu$s, the simulations presented here could be expected to be roughly $10^5/500 \times 75$\,s = 15,000 CPU-s, somewhat greater than the 4,000 CPU-s total simulation time for Models 3--5 to compute out to 100\,$\mu$s as reported in Table \ref{tab:model-cost-chi}. Our method would take roughly 700 CPU-s to calculate out to 10\,$\mu$s. However, these differences are in line with processor variations and it is reasonable to say the computational cost is comparable.

Where our new method shows significant advantage is in the scaling to many primaries with interacting radiolytic products since the initial set-up computational cost scales linearly while the time to evolve the chemistry is roughly constant (see Fig.~\ref{fig:model-3-comp-times}). Na\"ive SBS Monte Carlo computational times for the chemical stage can be expected to scale with the square of the number of reactive species since reactions between all possible partners must be considered. Hence, the scaled computational time for Monte Carlo to simulate the 10,000 Gy dose deposition would be $10^4$ times greater (i.e.~10,000 Gy/100 Gy raised to the second power), about $1.5 \times 10 ^ 8$ CPU-s or approximately 5 CPU-years. In contrast, our approach gives these results in about $\frac{1}{2}$ a CPU-day. There are many-body algorithms which can overcome this quadratic scaling (e.g.~\cite{Dehnen2002_HierarchicalON}) although these do not yet appear to have been implemented in SBS radiation chemistry computational codes.

The IRT method circumvents per–step pair searching by sampling first–passage times and updating an event queue, often achieving near–linear scaling in the number of reaction events~\cite{pimblott_laverne_jpca_1997,ramosmendez_irt_geant4_2020}. However, heavy rescheduling in dense spurs can erode this advantage~\cite{clifford_green_pilling_faraday_1986}. Full track–structure Monte Carlo adds further cost by first generating cascades of secondary electrons and excitations, yielding tens–to–hundreds of thousands of species for the chemical stage~\cite{friedland_partrac_2011,plante_cucinotta_hze_2008,uehara_monte_2006}. These challenges—particularly acute for high linear energy transfer (LET) tracks—have motivated growing interest in on–lattice RDME solvers and continuum reaction–diffusion PDEs, which scale with grid points and species rather than molecules, trading stochastic detail for tractability at long times and large system sizes~\cite{tran_geant4-dna_2021} although convergence properties in terms of the number of compartments used are yet to be established for these methods. In contrast, the method we present here automatically selects spatial and time discretization by error estimation (by a standard time-step doubling method) and does not involve complex book-keeping \cite{bradshaw_new_2023, bradshaw2025spectral}.

The new ultra-high dose deposition results this new approach has facilitated also warrant some discussion. As shown in Fig.~\ref{fig:model-3-dose-rates}, the G-value dynamics is roughly independent of dose (and hence reaction packet density) up to about 0.1\,ns. This phase of the dynamics is governed by the intra-packet dynamics with the reaction packets acting as if in isolation. However, the G-value dynamics begins to diverge after this with the highest density leading to more rapid depletion of \OH, \eaq \;and \HthreeOplus \; along with a commensurate increase in \Htwo \; and \HtwoOtwo \; production. The 100\,$\mu$s convergence of the G-values regardless of dose rate is a consequence of all reactions having reached completion with the asymptotic yields being reached. However, it is interesting to note that even at the highest dose (10,000 Gy), the G-value dynamics is distinct from that of the homogenous system. For example, the G-value for \eaq \; takes over 100 ns to halve in value for the homogenous calculation, whereas it happens in under 10 ns for 10,000 Gy deposition. 

The comparative rates of removal of \Hdot, \; \OH \; and  \eaq \; for the 10,000 Gy, 100 Gy inhomogeneous and 100 Gy homogenous simulations can be understood from the density of the reaction vertices and the spatial extent of species about these vertices. Each vertex emits an electron distribution and reactive species distributions which then further spread out due to diffusion. As the dose increases, so does the density of vertices. This means that the time for separate packets to interact decreases. Hence, the common G-value evolution seen out to about 0.1 ns can be ascribed to intra-packet reactions, i.e. interactions only between species originating from the same ionization/excitation vertex. At longer times, the curves for the different dose depositions separate the; inter-vertex reactions dominate more rapidly at higher dose depositions.

Since the dominant loss reactions of $\mathrm{e_{aq}^{-}}$ are bimolecular, 
their rates scale with the average of the product of the concentrations rather than the product of mean concentrations. 
High local overlap at 10,000~Gy therefore produces much larger local reaction rates, greatly accelerating $\mathrm{e_{aq}^{-}}$ loss. 
The homogeneous model, lacking both the high local densities of the 10,000~Gy case and the spatial isolation of 100~Gy, gives an intermediate decay timescale. 

These results point to the applicability of this model to ultra-fast laser driven radiolysis \cite{dromey_picosecond_2016,prasselsperger_realtime_2021} where related effects have been observed. These results also point to the applicability of the method to model electron track effects in nanoparticle dose enhancement \cite{McMahon2011Biological}. These high dose rate results also point to the need for caution when using homogenous models for radiation chemistry. Because the reaction vertices occur stochastically there can be significant various in vertex density across the irradiated region, a factor which is not captured in homogeneous models at all. 

In the near future we plan to publish a comparison between our model and G-values from fast ions. There are sets of experimental data, some of which have not been used to fine-tune Monte Carlo calculations and which are expected to produce noticeable differences \cite{Yamashita2008WaterRadiolysis2}. The computational scaling advantage of our new method will be expected to be most beneficial for high-LET due to the large numbers of interacting chemical species.

One of the advances of this new hybrid approach is that it is amenable to supporting a fully \emph{ab-initio} treatment, e.g.~through time dependent density functional theory calculations (TDDFT). The forms of the reaction packets for excitation and ionisation could be derived from very localised TDDFT calculations, concerning only a few nm$^3$ and over 1 ps. These forms would then be upscaled across greater distances and over greater times using the approach described here. This approach would be applicable to any liquid.

It is important to note that no new additional adjustable parameters have been used in formulating the models presented here. However, some parameters have been re-purposed from previous Monte Carlo calculations. For example, the pre-chemical branching ratios for \HtwoOstar \;were set by Kreipl et al.~\cite{kreipl_time_2009} to reproduce ps G-values to later be inherited into Geant4-DNA \cite{Shin2021Geant4DNAprechem} and thus into the calculations of Tran et al.~\cite{tran_geant4-dna_2021} and also used here. Although the tuning was done against ps ion-driven yields, the electron- and ion-physics is very similar. Accordingly, it is possible that the fine-tuning introduced a model-type dependence which then impacts on the Monte Carlo electron G-value agreement, i.e.~part of the agreement is a reflection of the model being tuned to experimental data. Similarly there are a set of RMS displacements used \cite{kreipl_time_2009, RamosMendez2021PMB} which were also adjusted to match pulsed radiolysis measurements of Wang et al.~\cite{wang_time-dependent_2018}, and again inherited by Tran et al.~\cite{tran_geant4-dna_2021}. Once again, possibly this process has added a model-dependent parametrisation. Historically, Monte Carlo has tended to be accepted as the `ground truth' but this is largely because there has been no competitor method. It is possible that this new approach will yield results closer to the experimental `ground truth' and will certainly provide the opportunity for methodological comparisons, particularly if \emph{ab-initio} parameterisation of the reaction packets is achieved. 

\section{Conclusion} \label{sec:conclusion}

A new method for calculating radiolytic effects of electron track structure, the fundamental building block of radiation chemistry, has been introduced. The method produces results comparable to those from Monte Carlo calculations although with superior  CPU-usage scaling at high numbers of chemical species. These superior scaling properties have been used to perform calculations of water radiolysis at unprecedentedly high dose depositions. However, the new method has been shown to over-estimate production of molecular species \Htwo \;and \HtwoOtwo \;due to `self-interactions'  within individual reaction packets. A correction term has been derived for this effect. Furthermore, the new method is amenable to taking input from short time, local \emph{ab initio} simulations, opening a pathway to long-time, large scale \emph{ab initio} simulations.

\section*{Author contributions}
All authors contributed equally to the conceptualization and authoring of the manuscript. MW exclusively wrote the core MIRaCLE code. CP and FC wrote the example-specific code which called the core code.

\section*{Conflicts of interest}
There are no conflicts to declare.

\section*{Data availability}

Data for this article, including Julia simulation and python plotting codes, numerical results and the figures are available on Figshare:
\href{https://figshare.manchester.ac.uk/account/articles/30940346}{https://figshare.manchester.ac.uk/account/articles/30940346}.

\section*{Acknowledgements}

CP thanks the Dalton Cumbrian Facility engagement budget and the University of Manchester Student Experience Internship (SEI) programme for providing summer studentships. MW thanks the Polish National Science Centre (SONATA-BIS-9), project no. 2019/34/E/ST1/00390, for the funding that supported some of this research.

We thank H.N.~Tran for useful discussions and for sharing data from \cite{tran_geant4-dna_2021}.

We are grateful for scientific input on research from Helen Steele (Sellafield), and to Jacob Anderson (UKNNL and UoM), Daniel Conn (UoM), Saoirse Currell (UoM), Zhanyue Leo Gao (UoM), Gus Taylor (UoM) and Thomas Wilkins (UKHSA and UoM) for contributing to the collaborative environment in the MIRaCLE Summer School of 2025. We thank Ian Hinder and Mike Jones (Research IT, UoM) for IT Support.





\bibliography{references} 
\bibliographystyle{rsc} 

\clearpage

\renewcommand{\thesection}{\Alph{section}}  
\setcounter{section}{0}


\clearpage              

\twocolumn[             
  \section*{Electronic Supplementary Information (ESI)}
  \addcontentsline{toc}{section}{Supplementary Information}
]

\renewcommand{\thefigure}{S\arabic{figure}}  
\setcounter{figure}{0}
\renewcommand{\thetable}{S\arabic{table}}  
\setcounter{table}{0}
\renewcommand{\theequation}{S\arabic{equation}}
\setcounter{equation}{0}

\renewcommand{\thesection}{S\arabic{section}}
\renewcommand{\thesubsection}{\thesection.\arabic{subsection}}
\renewcommand{\thesubsubsection}{\thesubsection.\arabic{subsubsection}}
\setcounter{section}{1}

Reference numbers refer to references in the main text.

The parameters used to develop the models have all been drawn from the scientific literature. The parameters defining the chemistry, in overview, are given in Table \ref{tab:combined-physchem-data}. Other parameters, governing more detailed aspects of the models developed are given in Tables \ref{tab:SI_vertex-probabilities}, \ref{tab:SI_ballarini-branching-ratios}, \ref{tab:SI_kreipl-branching-ratios}, \ref{tab:SI_widths} and \ref{tab:SI_vertex-mappings}.

\begin{table}[h]
\centering
\small
\begin{tabular}{lccc}
\toprule
\multicolumn{4}{l}{\textbf{Diffusion coefficients}~\cite{kreipl_time_2009}} \\
\midrule
Species &
\shortstack{Diffusion coefficient\\(\si{nm^2.ns^{-1}})} &
& \\
\OH           & 2.8 & & \\
\esol         & 4.9 & & \\
\HtwoOtwo     & 2.3 & & \\
\Htwo         & 4.8 & & \\
\Hdot         & 7.0 & & \\
\HthreeOplus  & 9.0 & & \\
\OHminus      & 5.0 & & \\
\midrule
\multicolumn{4}{l}{\textbf{Reaction network}~\cite{tran_review_2024}} \\
\midrule
\multicolumn{3}{l}{Reaction} &
\shortstack{Rate $k$\\(\si{M^{-1}.ns^{-1}})} \\
\multicolumn{3}{l}{$2$\esol{} + $2$\HtwoO{} $\rightarrow$ $2$\OHminus{} + \Htwo}                       & 6.36 \\
\multicolumn{3}{l}{\esol{} + \OH{} $\rightarrow$ \OHminus}                                              & 29.5 \\
\multicolumn{3}{l}{\esol{} + \Hdot{} + \HtwoO{} $\rightarrow$ \Htwo{} + \OHminus}                      & 25.0 \\
\multicolumn{3}{l}{\esol{} + \HthreeOplus{} $\rightarrow$ \Hdot{} + \HtwoO}                            & 21.1 \\
\multicolumn{3}{l}{\esol{} + \HtwoOtwo{} $\rightarrow$ \OH{} + \OHminus}                               & 11.0 \\
\multicolumn{3}{l}{$2$\OH{} $\rightarrow$ \HtwoOtwo}                                                    & 5.5  \\
\multicolumn{3}{l}{\Hdot{} + \OH{} $\rightarrow$ \HtwoO}                                             & 15.5 \\
\multicolumn{3}{l}{$2$\Hdot{} $\rightarrow$ \Htwo}                                                      & 5.03 \\
\multicolumn{3}{l}{\HthreeOplus{} + \OHminus{} $\rightarrow$ $2$\HtwoO}                                & 113  \\
\midrule
\multicolumn{4}{l}{\textbf{Radiolysis yields and initial homogeneous concentrations at $1$ps}~\cite{tran_geant4-dna_2021}} \\
\midrule
Species &
\shortstack{$G$-value\\(molecules/$100$eV)} &
\shortstack{Yield for $5$keV\\(particles)}  \\
\OH          & 5.24   & 262   \\
\esol        & 3.85   & 193   \\
\HtwoOtwo    & 0.0490 & 2.45   \\
\Hdot        & 0.73   & 36.5   \\
\HthreeOplus & 4.13   & 207   \\
\OHminus     & 0.0925 & 4.63  \\
\Htwo        & 0.2959 & 14.8  \\
\bottomrule
\end{tabular}
\caption{Summary of diffusion coefficients, reaction rate constants, G-values and the deduced initial radiolysis yields for 100 Gy dose deposition used in this work. Literature sources are indicated for each subsection header. Particle yields were determined by $N_\text{particles}=\mathrm{LET}\times E_\mathrm{dep}$.}
\label{tab:combined-physchem-data}
\end{table}
\begin{table}[H]
\centering
\small
\begin{tabular}{ll}
\toprule
\textbf{Vertex type} & \textbf{Probability (\%)} \\
\midrule
\HthreeOplus + \OH + \eaq & $86.7$ \\
\OH + \Hdot & $10.7$ \\
\Htwo + $^\bullet \mathrm{O}^\bullet$ & $02.6$ \\
\bottomrule
\end{tabular}
\caption{Assigned inelastic vertex types and probabilities calculated from \cite{kreipl_time_2009, ballarini_stochastic_2000} adjusting for vertices that don't produce radiolytic species.} \label{tab:SI_vertex-probabilities}
\end{table}

\vfill

\begin{table}[H]
    \centering
    \small
    \begin{tabular}{lc}
        \toprule
        \textbf{Component} & \textbf{1 MeV Electron} \\
        \midrule
        e--sub (\%)        & 37.1 \\
        H$_2$O$^+$ (\%)    & 37.1 \\
        A$^1$B$^1$ (\%)    & 8.7  \\
        B$^1$A$^1$ (\%)    & 9.1  \\
        Ry/db/de (\%)      & 7.5  \\
        \bottomrule
    \end{tabular}
    \caption{Percentage yield of excitation/ionisation products for $1$ MeV incident electrons from \cite{ballarini_stochastic_2000} at $1$ ps after irradiation.}
    \label{tab:SI_ballarini-branching-ratios}
\end{table}

\begin{table}[H]
\centering
\small
\renewcommand{\arraystretch}{1.2}
\begin{tabular}{>{\raggedright}p{2.7cm} >{\raggedright}p{3.6cm} >{\centering\arraybackslash}p{1cm}}
\toprule

\textbf{Process} & \textbf{Decay channel} & \textbf{Fraction (\%)} \\
\midrule
Ionisation (\HtwoOplus) &  & \\
\quad 1b$_1$, 3a$_1$, 1b$_2$, 2a$_1$, K & \HthreeOplus + \OH & 100 \\
\addlinespace
Excitation (\HtwoO$^*$) &  & \\
\quad A$^1$B$^1$ & \OH + \Hdot & 65 \\
& Relaxation to \HtwoO & 35 \\
\addlinespace
\quad B$^1$A$^1$ & \HthreeOplus + \OH + \eaq (auto-ionisation) & 55 \\
& \Htwo + $^\bullet\mathrm{O}^\bullet$ & 15 \\
& Relaxation to \HtwoO & 30 \\
\addlinespace
\quad Rydberg/diffuse & \HthreeOplus + \OH + \eaq & 50 \\
& Relaxation to \Htwo & 50 \\
\bottomrule
\end{tabular}
\caption{Decay channels and branching ratios from \cite{kreipl_time_2009}.} \label{tab:SI_kreipl-branching-ratios}
\end{table}

\begin{table}[H]
\centering
\small
\begin{tabular}{ll}
\toprule
\textbf{Species} & \textbf{Width}, $\mathrm{nm}$ \\
\midrule
\OH          & $1.10$ \\
\esol        & $12.1$ ($r_0$)\\ 
\HtwoOtwo    & $0.435$ \\
\Htwo        & $0.154$ \\
\Hdot        & $1.35$ \\
\HthreeOplus & $1.20$ \\
\OHminus     & $0.663$ \\
\bottomrule
\end{tabular}
\caption{Widths of the concentration profiles about a single event vertex. Widths are equivalent to Gaussian $\sigma$ for all species except for \esol. In this case, the quoted value represents the scale parameter, $r_0$, of a gamma distribution, with the form given in eqn. \ref{eqn:kreipl-gamma} of the main text. }
\label{tab:SI_widths}
\end{table}

\begin{table}[h]
\centering
\small
\begin{tabular}{ll}
\toprule
\textbf{Species} & \textbf{Vertex types} \\
\midrule
\HthreeOplus & \HthreeOplus + \OH + \eaq \\
\OH          & \HthreeOplus + \OH + \eaq \, and \,  \OH + \Hdot \\
\eaq         & \HthreeOplus + \OH + \eaq \\
\Hdot        & \OH + \Hdot \\
\Htwo        & \Htwo  + $^\bullet\mathrm{O}^\bullet$ \\
\HtwoOtwo    & \Htwo  + $^\bullet\mathrm{O}^\bullet$ \\
\OHminus     & \HthreeOplus + \OH + \eaq \\
\bottomrule
\end{tabular}
\caption{Mapping from species to vertex types from Table \ref{tab:SI_vertex-probabilities}. The left column shows the species which are placed at the vertex types in the right column. \HtwoOtwo \; is centred on the \Htwo + $^\bullet\mathrm{O}^\bullet$ location since the $^\bullet\mathrm{O}^\bullet$ reacts with bulk water producing two \OH \; radicals, which immediately react to form \HtwoOtwo \cite{kreipl_time_2009}. \OHminus is centred on ionisation vertices due to their formation by reactions between \OH \; and \eaq, noting that \OHminus cannot be produced directly by either of the other two vertex types.}\label{tab:SI_vertex-mappings}
\end{table}

\subsection{Model 2 and Widths for Cylindrical Gaussians}
\label{sec:SI_Model2}
In Model 2, the radiolytic species produced by the passage of a single electron are distributed isotropically along its path, which is treated as a straight line since there is negligible energy loss or scattering under these conditions \cite{bethe_molieres_1953, nist_star_2017}, i.e.~an electron-specific z-axis is created. The cylindrically symmetric distribution about this axis was taken to be a Gaussian probability density function (PDF) per unit volume of finding a dissociation fragment at a distance $r$ from this axis. In the work of \cite{kreipl_time_2009}, species were displaced from a collision vertex by a radial distance sampled from the following Gaussian radial PDF
\begin{equation} \label{eqn:kreipl-gaussian}
    f_{\mathrm{rad}}(r) = \sqrt{\frac{2}{\pi}}\frac{1}{\sigma^3}r^2\exp\bigg(-\frac{1}{2}\frac{r^2}{\sigma^2}\bigg),
\end{equation}
where the standard deviation, $\sigma$, was calculated from a set of root-mean-square (RMS) displacements of the species from the collision vertices, $r_\text{rms}=\sqrt{3}\sigma$. Further dissociation-channel specific displacements were made \cite{kreipl_time_2009}. The specific form of (\ref{eqn:kreipl-gaussian}), used in this work is
\begin{equation} \label{eqn:cylindrical-gaussian}
    f_{\mathrm{vol}}(r) = \frac{1}{2\pi\sigma^2L}\exp\bigg(-\frac{1}{2}\frac{r^2}{\sigma^2}\bigg),
\end{equation}
where $L$ is the length of the domain in the $z$ direction. (\ref{eqn:cylindrical-gaussian}) is the cylindrical analogue of (\ref{eqn:kreipl-gaussian}), however where (\ref{eqn:kreipl-gaussian}) represents the probability density of locating a particle at a radius $r$ in $[\mathrm{nm}^{-1}]$, (\ref{eqn:cylindrical-gaussian}) represents the probability density of locating a particle at $(r,z)$ in $[\mathrm{nm}^{-3}]$. 

Three dissociation processes were considered \cite{kreipl_time_2009}, with the spatial distributions of all species given by Gaussians. Each process is discussed in turn below. The choice of Gaussians is appropriate since it is the exact continuum representation of the Monte Carlo dissociation processes where there is only one production mechanism for a given species. Here, we associated a single Gaussian PDF for any species created in an ionisation/excitation event. Accordingly, one width parameter, $\sigma_i$, per species encodes the spatial inhomogeneity due all the dissociation processes able to produce that particle. The width parameter takes into account the ballistics for all production mechanisms relevant to a given species.

In the dissociation process 
\[
\HtwoOplus + \HtwoO \rightarrow \HthreeOplus + \OH,
\]
the \HtwoOplus \; is displaced by $r_\text{rms}=2.0 \; \mathrm{nm}$ due to a series of fast charge transfers, and then two channels with equal probability form either \HthreeOplus \;or \OH \;displaced by $r_\text{rms}=0.8\mathrm{nm}$ from the \HtwoOplus \;position, during the dissociation process. The other species remains at the previous location of the \HtwoOplus \cite{kreipl_time_2009}. Since each step corresponds to sampling a Gaussian PDF, the probability density of finding a specific dissociation fragment after two steps is the convolution of the two composite Gaussians with width $\sigma = \sqrt{\sigma_1^2 + \sigma_2^2}$. 

Since generally the sum of two Gaussians is not a Gaussian, to calculate the standard deviation of a single Gaussian to approximate the PDF associated with the sum of the competing pathways, the weighted mean of the standard deviations was used. In the dissociation process under discussion, the widths of the single Gaussian approximations for each species is $\sigma=p_1\sigma_1+p_2\sqrt{\sigma_1^2+\sigma_2^2}$, where $p_1=p_2=0.5$ are the probabilities of each pathway, $\sigma_1=\frac{2.0}{\sqrt{3}}\mathrm{nm}$ and $\sigma_2=\frac{0.8}{\sqrt{3}}\mathrm{nm}$ are the widths associated with the RMS displacements of each step.

In the dissociation process
\[
\HtwoO^*\rightarrow\OH+\Hdot,
\]
the RMS displacement used by Monte Carlo studies is $2.4\mathrm{nm}$, and the \OH \; and \Hdot \; are displaced by $\frac{-1}{18}r$ and $\frac{+17}{18}r$ (due to conservation of momentum), respectively, where $r$ is the sampled distance from equation \ref{eqn:kreipl-gaussian} \cite{kreipl_time_2009}. In this scheme, the  widths associated with the Gaussians are $\sqrt{1/18}\sigma_0$ and $\sqrt{17/18}\sigma_0$, respectively for  \OH \; and \Hdot, where $\sigma_0=\frac{2.4}{\sqrt{3}}\mathrm{nm}$. 

Finally, in the dissociation process 
\[
\HtwoO^* \rightarrow \Htwo \mathrm{+} ^\bullet\mathrm{O}^\bullet \rightarrow \Htwo + \OH + \OH \rightarrow \Htwo + \HtwoOtwo,
\]
the oxygen atom forms two \OH \;radicals rapidly in a reaction with water \cite{kreipl_time_2009}, both of which then immediately react to form \HtwoOtwo  \cite{tran_geant4-dna_2021}. The RMS distance associated with the first step is $r_{\text{rms}}=0.8\mathrm{nm}$, with the \Htwo \ and $^\bullet \mathrm{O}^\bullet$ dislocating by $\frac{-2}{18}r$ and $\frac{+16}{18}r$, respectively \cite{kreipl_time_2009}.  The location of \HtwoOtwo \;formation is treated as being at the same site as $^\bullet\mathrm{O}^\bullet$ formation. Therefore, the widths associated with this dissociation channel for \Htwo \;and \HtwoOtwo \; are $\sqrt{2/18}\sigma_0$ and $\sqrt{16/18}\sigma_0$, respectively, where $\sigma_0=\frac{0.8}{\sqrt{3}}\mathrm{nm}$.

Since \OH \;is a final product in two of the three dissociation processes, the final width for this species was taken as the weighted mean of the two constituent widths in each of the relevant dissociation pathways, with the weights given by the branching ratios for each pathway, see Table \ref{tab:SI_vertex-probabilities}. In all other cases, the widths were as described above.

The branching ratios presented in  Table \ref{tab:SI_vertex-probabilities} were calculated using the branching ratios for the final products after the physical stage of irradiation, shown in Table \ref{tab:SI_ballarini-branching-ratios} and the final products after the physicochemical stage, shown in Table \ref{tab:SI_kreipl-branching-ratios}. The values in Table \ref{tab:SI_vertex-probabilities} were calculated by summing up the total probabilities that a given inelastic collision vertex for a $1 \; \mathrm{MeV}$ incident electron would produce a given final set of reactants, following a decision tree constructed from all the pathways. Some physicochemical pathways result in `production' of \HtwoO \;due to a relaxation process. Since water is in excess, these \HtwoO \;production event vertices can be ignored. Hence, all vertices in our model are of the three types listed in Table \ref{tab:SI_vertex-probabilities}. However, the energy deposited in the water as a result of the relaxation processes is still considered.

The final set of Gaussian widths used in Models 2--5 is shown in Table \ref{tab:SI_widths}. \OHminus \;is a product of chemical reactions after the physicochemical stage. It is produced by reactions between solvated electrons and \OH, \HtwoOtwo, \Hdot, and \HtwoO. Consequently, the spatial distribution of \OHminus \;will be dictated by the spatial distribution of these reactants. Since solvated electrons form only at sites of ionisation collisions, where only \HthreeOplus \;and \OH \;are produced, the initial \OHminus \;yield will be dominated by reactions between \eaq \;and \OH. Therefore, the initial spatial distribution of \OHminus \;will be proportional to $k_2\rho_\mathrm{e^-_{aq}}(\bm{x})\rho_{\mathrm{OH}}(\bm{x})$. The width of \OHminus was calculated by performing a least-squares fit to the product of the two reactant distributions $\rho_\mathrm{e^-_{aq}}(\bm{x})\rho_{\mathrm{OH}}(\bm{x})$, using a Gaussian of width $\sigma$. $\sigma$ was the fitting parameter with the fit being subject to the constraint that the target Gaussian distribution has the same integral as that of the product of the reactant distributions. 

In Model 2, electrons were also distributed following a Gaussian distribution, though this is not the case in later models. Monte Carlo studies have shown that the distribution of thermalisation distances for secondary electrons may follow Gaussian or modified exponential forms \cite{terrissol_simulation_1990}. In both cases, the width is dependent on the mean thermalisation distance $r_0$, and is given by $\sigma=\frac{2\sqrt{2}}{\sqrt{\pi}}r_0$. The dependence of mean thermalisation distance on energy has been explored extensively in the literature across a wide range of energies \cite{meesungnoen_low-energy_2002, plante_cross_2009}. 

Here, $r_0$ was taken to be the mean thermalisation distance for the mean secondary electron energy as produced by $1\mathrm{MeV}$ incident electrons in water, as calculated from Figure 9 of \cite{uehara_cross-sections_1993} by averaging over $10^7$ random samples of the secondary energy. The mean secondary electron energy $E_0$ was found to be $E_0=48.2\mathrm{eV}$. Using the relativistic calculation in Figure 11 of \cite{plante_cross_2009}, the mean thermalisation distance for a $48.2\mathrm{eV}$ electron is $r_0=12.1\mathrm{nm}$, and so the width of the Gaussian approximation is $\sigma=7.59\mathrm{nm}$. 

In Model 2, radiolytic species were distributed about $N$ straight electron tracks, parallel with the $z$ axis. The positions of the tracks were randomly sampled from a uniform distribution in $[0,200]\; \mathrm{nm}$ for $x$ and $y$. The average energy lost by a $1 \; \mathrm{MeV}$ electron traversing $200\mathrm{nm}$ is approximately $\sim40 \; \mathrm{eV}$ according to NIST-ESTAR stopping power data \cite{nist_star_2017}. Since this is on the order of the mean ionisation energy per inelastic vertex, $I=79.9 \; \mathrm{eV}$ \cite{hiraoka_energy_1994}, the number of electron tracks was calculated as $N=\frac{E_{\mathrm{dep}}}{I}\eta$ where $\eta$ is the mean inelastic radiolysis fraction, the average fraction of inelastic collisions that do not undergo relaxation to \HtwoO. $\eta$ was calculated from the branching ratios in Table \ref{tab:SI_ballarini-branching-ratios} by totalling the probability that a given inelastic collision results in relaxation to water using the associated decision tree. The fraction was found to be $\eta=0.848$. $N$ was rounded to the nearest integer value.

The full concentration field for each species was then the normalised sum of the cylindrical Gaussians over all incident electron coordinates (the probability density function for a single particle of a given species in the world), multiplied by the total number of species in the world, as given in Table \ref{tab:combined-physchem-data}. Explicitly, the concentrations were
\begin{equation} \label{eqn:cylindrical-concentration}
    \rho_i(x,y,z) = \frac{M_i}{2\pi \sigma_i^2 L} \frac{10^{24}}{N_A} \frac{1}{N}\sum_{j}^N\exp\bigg(-\frac{1}{2}\frac{(x-x_j)^2+(y-y_j)^2}{\sigma_i^2}\bigg)
\end{equation}
where $M_i$ is the total number of particles of species $i$, in the volume ($M_i=\frac{E_{\mathrm{dep}}}{100}g_i$ for a g-value $g_i$), $x_j$ and $y_j$ are the coordinates of the incident electrons, and $\sigma_i$ are the widths of the Gaussians for each species.

\subsection{Generation of Spherical Gaussians Distributions in Model 3}

Models 3--5 use a regularised gamma distribution, motivated by the approximations used by \cite{kreipl_time_2009}. In the gamma distribution distribution, $r_0$ directly sets the scale of the distribution instead of $\sigma$. All of the same physical processes were treated in the same way as for Model 2 although for species other than the solvated electron spherical Gaussians were used throughout.  

The generation of the $x$ and $y$ coordinates for the collision vertices was the same as used in Model 2 as was the calculation of the number of collisions that induce radiolytic chemistry, $N$. The generation of points in $z$ at which to place Gaussians along an electron trajectory followed a typical Monte Carlo scheme. Starting from an offset position of $1 \; \mathrm{\mu m}$ in $z$ in front of the simulation box, successive $z$ positions were obtained by sampling step lengths from an exponential distribution with mean $r_{\mathrm{mfp}}=320 \; \mathrm{nm}$, corresponding to the inelastic mean free path calculated using ionisation and excitation cross sections from \cite{plante_cross_2009}. Each sampled step was added to the current depth until the upper boundary of the box was reached. Whenever a sampled position lay inside the simulation volume (between $z_\mathrm{min}$ and $z_\mathrm{max}$), it was used as a collision site.  New tracks were generated with this procedure until the total number of vertices was greater than or equal to $N$. In $50\%$ of cases, the last track was discarded to keep the average number of vertices equal to $N$. This procedure generated columns of interaction points along fixed $(x,y)$ coordinates, consistent with the expected exponentially distributed inter-collisional distances. Labelling the collision vertices $(x_j, y_j,z_j)$, the full concentration field for all species $i$ except electrons was
\begin{equation} \label{eqn:spherical-concentration}
    \begin{split}
    \rho_{i\ne e^-_{\mathrm{aq}}}(x,y,z) &= \frac{M_i}{(2\pi \sigma_i^2)^{3/2}} \frac{1\times10^{24}}{N_A} \times \\ &\frac{1}{N}\sum_{j}^N\exp\bigg(-\frac{1}{2}\frac{(x-x_j)^2+(y-y_j)^2+(z-z_j)^2}{\sigma_i^2}\bigg),
    \end{split}
\end{equation}
following the conversion of equation \ref{eqn:cylindrical-concentration} from cylindrical to spherical symmetry about the collision vertices. Similarly, the concentration of solvated electrons was given by
\begin{equation} \label{eqn:spherical-concentration}
\begin{split}
    \rho_{e^-_{\mathrm{aq}}}(x,y,z) &= \frac{M_{e^-_{\mathrm{aq}}}}{\pi r_0^2} \frac{1\times10^{24}}{N_A} \times \\ &\frac{1}{N}\sum_{j}^N
        \frac{r_j(x,y,z)-r_{\mathrm{cut}}}{r_j^2(x,y,z)}\exp\Bigg(\frac{-2(r_j(x,y,z)-r_{\mathrm{cut}})}{r_0}\Bigg)
        \end{split}
\end{equation}
for $r_j(x,y,z)\ge r_{\mathrm{cut}}$, where $r_j(x,y,z) = (x-x_j)^2+(y-y_j)^2+(z-z_j)^2$ is the radial distance from the collision vertex $j$, and $\rho_{\mathrm{e^-_{aq}}}=0$ otherwise.

In this model, the value of $r_0$ used is the mean thermalisation distance, $r_0 = 12.1\mathrm{nm}$, for a $48.2\mathrm{eV}$ electron, as calculated for Model 2. All other widths associated with the distribution of radiolytic species were kept the same as used in Model 2.

\subsubsection*{Model 4: Physical Stage Modelling} \label{sec:SI_method-model4}

This model uses the information of Table \ref{tab:SI_kreipl-branching-ratios}, with the ratio between ionisation and excitation from Table \ref{tab:SI_ballarini-branching-ratios}. The calculated probabilities were taken from Table \ref{tab:SI_vertex-probabilities}. Gaussian distributions were once more centered on these vertices.

After the vertices were generated exactly as in Model 3, they were additionally labelled with a vertex type stochastically according to the probabilities given in Table \ref{tab:SI_vertex-probabilities}. The label denotes which physicochemical products are produced at the vertex. The concentration of a given species then depends on the vertex types at which the species is produced, governed by \emph{vertex mappings} given in Table \ref{tab:SI_vertex-mappings}. Some mappings are obvious, because the species are a direct product of the dissociation pathway associated with the vertex type. \OHminus \;has been mapped to the ionisation vertex, \HthreeOplus + \OH + \eaq, since this is the only vertex type that has solvated electrons present, which are necessary for the production of \OHminus, as discussed in the context of the \OHminus \;spatial distribution earlier. \HtwoOtwo \;is mapped to the \Htwo  + $^\bullet\mathrm{O}^\bullet$ vertex due to the $^\bullet\mathrm{O}^\bullet + \HtwoO \rightarrow 2\OH \rightarrow \HtwoOtwo$ mechanism \cite{tran_geant4-dna_2021}. 

The concentration fields then follow equation \ref{eqn:spherical-concentration} different mixtures of reaction species are centred about different vertices, $(x_{j},y_{j},z_{j})$.

\subsubsection*{Model 5: Physical and Physicochemical Stage Modelling} \label{sec:SI_method-model-5}

Model 5 includes the effect of the ballistics from the dissociation displacing pairs of radiolytic species moving apart from their common parent molecular species. Practically, this displacement of vertices depends on the vertex type. The main process of interest is the dissociation of water in the A$^1$B$^1$ state to form  \OH + \Hdot, since these are highly reactive species. \Htwo + \HtwoOtwo \; was not considered since both species are stable. Because the species at the \HthreeOplus + \OH + \eaq \; vertex are formed via proton transfer $\HtwoOplus + \HtwoO \rightarrow \HthreeOplus + \OH$, modelling of the spatial distribution is not possible using simple kinematic considerations. As a result, dissociative separation of these species has been neglected also. Therefore, only the displacements of \OH \; and \Hdot \; from the A$^1$B$^1$ vertex were considered in this model.

The positions of the centre of the Gaussians for \OH \; and \Hdot \; were displaced by $-\frac{1}{18}r_{\mathrm{rms}}$ and $+\frac{17}{18}r_{\mathrm{rms}}$ in $z$ respectively, where $r_{\mathrm{rms}}$ is the RMS dissociation distance for the \OH + \Hdot \; dissociation, $r_{\mathrm{rms}}=2.4\mathrm{nm}$ as earlier.

\subsection{Pulse Radiolysis Details} \label{sec:SI_method-pulse-radiolysis}

Wang et al.~\cite{wang_time-dependent_2018} performed a picosecond pulse radiolysis study of $6$--$8 \; \mathrm{MeV}$ electrons into \HtwoO \; and determined the time dependent yields of \eaq \; and \OH \; from $0 \; \mathrm{ns}$ to $8 \; \mathrm{ns}$. The beam in the experiment entered the reaction cell with a Gaussian profile with a FWHM of  $4.6 \; \mathrm{mm}$ and $3.7 \; \mathrm{mm}$, for $x$ and $y$ respectively, and exited with a FWHM of $5.5\mathrm{mm}$ and $4.8\mathrm{mm}$, respectively. Hence, the current density, and the resultant density of ionisation/excitation vertices varies across the sample volume. 

The sample had an optical depth of $5 \; \mathrm{mm}$. The attenuation of the electron energy over this distance can was neglected; stopping power data \cite{nist_star_2017},  predicts about a $\sim100 \; \mathrm{keV}$ energy loss. The absorbed dose in the medium was $42.5\; \mathrm{Gy}$, and both the width and height of the sample were taken to be $4 \; \mathrm{mm}$ from Figure 3 of Wang et al.~\cite{wang_time-dependent_2018}. Although scattering is a significant effect at the scale of the sample, the electrons travel along straight lines in $z$ at the nanoscale.

A simulation with a $200\times200\times200\,\mathrm{nm^3}$ cube was performed with a dose chosen to approximate the average dose across the experimental sample for times from $1 \; \mathrm{ps}$ to $8 \; \mathrm{ns}$ using Model 3. This was achieved by locating the simulation box 1 standard deviation in both $x$ and $y$ away from the intensity maximum and half-way along the sample in the $z$ direction. Since the simulation volume is sufficiently small compared to the sample size and dose distribution, no modifications were needed to be made to the distributions from which the vertices were drawn. Boundary conditions were chosen to be Neumann at the boundary (reflective).

\subsection{Modified Rate Constants} \label{sec:SI_modified-const}

\paragraph{Derivation of the effective rate constant $k^{\mathrm{eff}}_{AA}$.}
We now justify the form
\[
k^{\mathrm{eff}}_{AA} \approx k_{AA}\,\frac{n_A}{n_A + J_{0,A}(t_{\mathrm{eff}})},
\]
by explicitly comparing the reaction rate obtained from the continuum packet model (with self-interactions) to the rate that would arise from a picture in which radicals from the same packet do not react with one another.

Consider a single species $A$ undergoing a diffusion-controlled homodimerisation $A{+}A\to\text{products}$ with microscopic rate constant $k_{AA}$.
In the continuum packet formulation, the concentration field of $A$ is written as a sum over $N_A$ identical packets,
\[
c_A(\mathbf r,t) \;=\; \sum_{i=1}^{N_A} \rho_i(\mathbf r,t),
\]
where each packet $\rho_i$ is centred on a vertex $\mathbf R_i$ and has the same normalized shape:
\[
\rho_i(\mathbf r,t) \;=\; \rho_A(\mathbf r-\mathbf R_i,t),
\qquad \int \rho_A(\mathbf r,t)\,d^3r = 1.
\]
The mean--field reaction rate density for a bimolecular homodimer is conventionally written as $\tfrac{1}{2}k_{AA}c_A^2$ (the factor $\tfrac{1}{2}$ avoids double counting of pairs), so the total reaction rate in the continuum model is
\[
R_{\mathrm{sim}}(t) \;=\; \frac{1}{2}k_{AA} \int c_A(\mathbf r,t)^2\,d^3r.
\]
Expanding $c_A^2$ gives
\[
\int c_A^2\,d^3r
= \sum_{i=1}^{N_A} \int \rho_i^2\,d^3r
+ 2\sum_{1\le i<j\le N_A} \int \rho_i(\mathbf r,t)\,\rho_j(\mathbf r,t)\,d^3r.
\]
The first term contains the (spurious) self-interactions of each packet with itself; the second term contains the physically relevant interactions between distinct packets.
We define the single-packet self-overlap
\[
J_{0,A}(t)
= \int \rho_A(\mathbf r,t)^2\,d^3r,
\]
so that $\int \rho_i^2\,d^3r = J_{0,A}(t)$ for all $i$.

For $i\neq j$ the overlap integral
\(
I_{ij}(t) = \int \rho_i \rho_j\,d^3r
\)
depends on the separation of the corresponding vertices.
To proceed analytically we adopt a mean-field approximation and assume a homogeneous, uncorrelated distribution of vertices in a volume $V$ (an idealized 3D Poisson process).
In that case, the ensemble average of $I_{ij}$ is
\[
\big\langle I_{ij}(t)\big\rangle
\approx \frac{1}{V},
\]
i.e.\ the average product of two independent, unit-normalized densities is the inverse of the volume.
Taking expectations and using the combinatorial identity
\(\sum_{1\le i<j\le N_A}1 = \tfrac{1}{2}N_A(N_A-1)\), we obtain
\begin{align*}
\Big\langle \int c_A^2\,d^3r \Big\rangle
&= \sum_{i=1}^{N_A} \big\langle \int \rho_i^2\,d^3r \big\rangle
+ 2\sum_{1\le i<j\le N_A} \big\langle I_{ij}(t)\big\rangle \\
&= N_A J_{0,A}(t)
+ 2 \cdot \frac{1}{2}N_A(N_A-1)\,\frac{1}{V} \\
&= N_A J_{0,A}(t) + \frac{N_A(N_A-1)}{V}.
\end{align*}
Consequently, the average simulated reaction rate is
\[
\big\langle R_{\mathrm{sim}}(t)\big\rangle
= \frac{1}{2}k_{AA}\,\Big\langle \int c_A^2\,d^3r \Big\rangle
= \frac{1}{2}k_{AA}\left[N_A J_{0,A}(t) + \frac{N_A(N_A-1)}{V}\right].
\]

In a physically correct picture where radicals from the same packet do not react with one another, only pairs from different packets should contribute.
If we regard each packet as representing one effective $A$-particle, then the number of distinct reactive pairs is $\binom{N_A}{2}=N_A(N_A-1)/2$, and the associated mean-field rate is
\[
R_{\mathrm{true}}(t)
\approx k_{AA}\,\frac{N_A(N_A-1)}{2V}.
\]
Here the factor $N_A(N_A-1)/2$ explicitly counts all unordered pairs.
We emphasise that this is already an approximation: we have assumed uniform mixing at scale $V$ and neglected any residual spatial correlations between packet centres beyond their mean density.

We now seek a single effective rate constant $k^{\mathrm{eff}}_{AA}$ such that
\[
\big\langle R_{\mathrm{true}}(t)\big\rangle
= \big\langle R_{\mathrm{sim}}^{\mathrm{(eff)}}(t)\big\rangle
= \frac{1}{2}k^{\mathrm{eff}}_{AA}\,\Big\langle \int c_A^2\,d^3r \Big\rangle,
\]
at a time relevant to determining the long-time G-values.
Equating the expressions and cancelling the common factor $1/2$ gives
\[
k_{AA}\,\frac{N_A(N_A-1)}{V}
= k^{\mathrm{eff}}_{AA}\left[N_A J_{0,A}(t) + \frac{N_A(N_A-1)}{V}\right].
\]
Solving for $k^{\mathrm{eff}}_{AA}$ yields the finite-$N_A$ expression
\[
k^{\mathrm{eff}}_{AA}(t)
= k_{AA}\,\frac{\dfrac{N_A(N_A-1)}{V}}{N_A J_{0,A}(t) + \dfrac{N_A(N_A-1)}{V}}
= k_{AA}\,\frac{\dfrac{N_A-1}{V}}{J_{0,A}(t) + \dfrac{N_A-1}{V}}.
\]
Introducing the packet number density $n_A = N_A/V$, this can be written as
\[
k^{\mathrm{eff}}_{AA}(t)
= k_{AA}\,\frac{n_A - \dfrac{1}{V}}{J_{0,A}(t) + n_A - \dfrac{1}{V}}.
\]
In the regime of interest, $N_A$ is large and the term $1/V$ is negligible compared to $n_A$, so that
\[
k^{\mathrm{eff}}_{AA}(t)
\approx k_{AA}\,\frac{n_A}{n_A + J_{0,A}(t)}.
\]
This is the form used in the main text, evaluated time $t=t_{\mathrm{eff}}$:
\[
k^{\mathrm{eff}}_{AA}(t_{\mathrm{eff}})
= k_{AA}\,\frac{n_A}{n_A + J_{0,A}(t_{\mathrm{eff}})}.
\]

Heuristically, $t_{\mathrm{eff}}$ should correspond to the point at which the initially localized radical packets have diffused far enough that the system has largely ``forgotten'' the initial localisation of species about reaction vertices, and further evolution is controlled mainly by large-scale mixing rather than by the artificial self-overlap within individual packets.

Recalling $N_A$ be the number of reaction vertices in a volume $V$, so that the vertex number density is
\[
n_A = \frac{N_A}{V}.
\]
For a statistically homogeneous distribution of vertices (e.g.\ an idealized 3D Poisson process), the typical distance between neighbouring vertices (the ``mean spacing'') scales as
\[
\ell_A \sim n_A^{-1/3}.
\]
This follows from the fact that any volume of order $\ell_A^3$ contains, on average, one vertex, i.e.\ $\ell_A^3 \sim 1/n_A$.

For a pair of radicals undergoing a diffusion-controlled homodimerisation $A{+}A\to\text{products}$,
if each radical has diffusion coefficient $D_A$, then the \emph{relative} coordinate $\mathbf r_{\mathrm{rel}}$ diffuses with coefficient
\[
D_{\mathrm{rel}} = 2D_A,
\]
and, by the Einstein relation in three dimensions, the mean--squared separation grows as
\[
\big\langle r_{\mathrm{rel}}^2(t) \big\rangle = 6 D_{\mathrm{rel}} t.
\]
We can therefore define a time-dependent characteristic diffusion length
\[
\ell_{\mathrm{diff}}(t) = \sqrt{\big\langle r_{\mathrm{rel}}^2(t) \big\rangle} = \sqrt{6 D_{\mathrm{rel}} t},
\]
which measures the typical change in separation of two radicals at time $t$.

At very early times, $\ell_{\mathrm{diff}}(t) \ll \ell_A$, and radicals remain confined within the neighbourhood of the vertex at which they were created.
In the continuum packet representation, this corresponds to a strong overlap of the same packet with itself, and hence an artificially enhanced self-reaction propensity.
At very late times, $\ell_{\mathrm{diff}}(t) \gg \ell_A$, and radicals have sampled regions encompassing many different vertices; individual packets have broadened and overlapped, and the local concentrations are effectively controlled by the coarse-grained mean density rather than by the initial vertex structure.
The transition between these two regimes occurs when the typical diffusion length becomes comparable to the vertex spacing, i.e.\ when
\[
\ell_{\mathrm{diff}}(t_{\mathrm{eff}}) \approx \ell_A.
\]
Substituting the expressions above and solving for $t_{\mathrm{eff}}$ gives
\[
\sqrt{6 D_{\mathrm{rel}} t_{\mathrm{eff}}} \approx \ell_A
\quad\Longrightarrow\quad
t_{\mathrm{eff}} \approx \frac{\ell_v^{2}}{6 D_{\mathrm{rel}}}
= \frac{n_A^{-2/3}}{6 D_{\mathrm{rel}}}.
\]
Thus $t_{\mathrm{eff}}$ is the time at which the relative diffusion length between reacting radicals matches the characteristic spacing between vertices.
From the point of view of correcting the self-interaction bias, this choice has a clear physical meaning: up to $t_{\mathrm{eff}}$, the overestimated self-overlap within each packet can significantly influence radical--radical encounter statistics, and beyond $t_{\mathrm{eff}}$ the system has become sufficiently mixed that the detailed packet origin of each radical is largely irrelevant.
Evaluating the packet self-overlap $J_{0,A}(t)$ and hence the effective rate constant $k^{\mathrm{eff}}_{AA}$ at this crossover time provides a single, dose- and geometry-specific correction that is tuned to the regime where the long-time (asymptotic) G-values are decided, while still being simple enough to implement as a fixed rate constant in the continuum model.

\paragraph{Assumptions and limitations.}
The derivation above relies on several simplifying assumptions:
(i) we assume a homogeneous, uncorrelated (Poisson) distribution of packet centres, so real track-structure correlations (clustering along tracks and within spurs) are not explicitly included;
(ii) we replace the true distribution of overlap integrals $I_{ij}$ by its mean value $1/V$, which is a mean-field approximation;
(iii) we take the large-$N_A$ limit so that terms of order $1/V$ can be neglected relative to $n_A$; and
(iv) we evaluate $J_{0,A}(t)$ at a single effective time $t_{\mathrm{eff}}$, thereby compressing a genuinely time-dependent correction into a single constant.
These approximations mean that $k^{\mathrm{eff}}_{AA}$ should be viewed as a dose- and geometry-specific renormalisation that is designed to reproduce long-time (asymptotic) G-values for a given irradiation regime, rather than as a universally valid replacement for the microscopic rate $k_{AA}$.
In particular, we do not expect a single $k^{\mathrm{eff}}_{AA}$ determined in this way to be transferable across very different doses or geometries, nor to reproduce early-time kinetics in detail.
Nonetheless, within these limitations, the construction provides a controlled and transparent way to compensate for the artificial self-interaction inherent in the  continuum packet representation.

We are currently conducting a series of computational experiments on the selection of $t_{\mathrm{eff}}$ and a generalization of this correction procedure. In particular, we note that the assumption of an uncorrelated (Poisson) distribution of packet centres is a restrictive one. The practical significance of this restriction will be the focus of a future publication.

\end{document}